\documentclass[pre,11pt,reprint,superscriptaddress]{revtex4-1}

\usepackage{amsmath}
\usepackage{graphicx}
\renewcommand{\vec}{\mathbf}
\newcommand{\mat}{\mathbf}
\newcommand{\Ord}{\mathrm{O}}
\newcommand{\Tr}{\mathop\mathrm{Tr}}

\begin{document}

\title{Message passing on networks with loops}
\author{George T. Cantwell}
\affiliation{Department of Physics, University of Michigan, Ann Arbor, Michigan, USA}
\author{M. E. J. Newman}
\affiliation{Department of Physics, University of Michigan, Ann Arbor, Michigan, USA}
\affiliation{Center for the Study of Complex Systems, University of Michigan, Ann Arbor, Michigan, USA}

\begin{abstract}
  In this paper we offer a solution to a long-standing problem in the study of networks.  Message passing is a fundamental technique for calculations on networks and graphs.  The first versions of the method appeared in the 1930s and over the decades it has been applied to a wide range of foundational problems in mathematics, physics, computer science, statistics, and machine learning, including Bayesian inference, spin models, coloring, satisfiability, graph partitioning, network epidemiology, and the calculation of matrix eigenvalues.  Despite its wide use, however, it has long been recognized that the method has a fundamental flaw: it only works on networks that are free of short loops.  Loops introduce correlations that cause the method to give inaccurate answers at best, and to fail completely in the worst cases.  Unfortunately, almost all real-world networks contain many short loops, which limits the usefulness of the message passing approach.  In this paper we demonstrate how to rectify this shortcoming and create message passing methods that work on any network.  We give two example applications, one to the percolation properties of networks and the other to the calculation of the spectra of sparse matrices.
\end{abstract}

\maketitle

\section{Introduction}
Networks occur in a wide range of contexts in physics, biology, computer science, engineering, statistics, the social sciences, and even arts and literature~\cite{Newman18c}.  Message passing, also known as belief propagation or the cavity method, is a fundamental technique for quantitative calculation of a wide range of network properties~\cite{Bethe35,Pearl82,MM09}, with applications to Bayesian inference~\cite{Pearl82}, NP-hard computational problems~\cite{MM09}, statistical physics~\cite{MM09,YGDM11,KNZ14}, epidemiology~\cite{KN10a}, community detection~\cite{DKMZ11a}, and signal processing~\cite{Gallager63,FM98}, among many other areas.  Message passing can be used both as a numerical method for performing explicit computer calculations and as a tool for analytic reasoning about network properties, leading to new formal results about percolation thresholds~\cite{KNZ14}, algorithm performance~\cite{DKMZ11a}, spin glasses~\cite{MPV87}, and other topics.  Many of the most powerful new results concerning networks in recent years have been derived from applications of message passing in one form or another.

Despite the central importance of the message passing method, however, it also has a substantial and widely discussed shortcoming: it only works on trees, i.e.,~networks that are free of loops~\cite{MM09}.  More generously, one could say that it works to a good approximation on networks that are ``locally tree-like,'' meaning that they contain only long loops but no short ones, so that local neighborhoods within the network take the form of trees.  However, most real-world networks, those that occur in practical applications of the method, contain many loops, including many short ones.  When applied to such ``loopy'' networks the method can give poor results, and in the worst cases can even fail to converge to an answer at all.

In this paper, we propose a remedy for this problem.  We present a series of methods of increasing elaboration for the solution of problems on networks with loops.  The first method in the series is equivalent to the standard message passing algorithm of previous work, which gives poor results in many cases.  The last in the series gives exact results on any network with any structure, but is too complicated for practical application in most situations.  In between lies a range of methods that give progressively better approximations, and which can be highly accurate in practice, as we will show, yet still simple enough for ready implementation.  Indeed even the second member of the series---just one step better than the standard message passing approach---already gives remarkably good results in real-world conditions.  We demonstrate the efficacy of our approach with two example applications.  The first is to the solution of the bond percolation problem on an arbitrary network, including the calculation of the size of the percolating cluster and the complete distribution of sizes of small clusters.  The second is to the calculation of the spectra of sparse symmetric matrices, where we show that our method is able to calculate the spectra of matrices far larger than those accessible by conventional numerical methods.

A number of approaches have been proposed previously for message passing on loopy networks.  The most basic, which goes by the name of ``loopy belief propagation,'' is simply to deploy the standard message passing equations, ignoring the fact that they are known to be incorrect in general.  While this might seem rash, it gives reasonable answers in some cases~\cite{FM98} and there are formal results showing that it can give bounds on the true value of a quantity in others~\cite{MM09,KNZ14}.  Perturbation theories that treat loopy belief propagation as a zeroth-order approximation have also been considered~\cite{chertkov2006loop}.  Broadly, it is found that these methods are suitable for networks that contain only a handful of short loops, but not for networks with many loops.

Some progress has been made for the case of networks that are composed of small subgraphs or ``motifs'' which are allowed to contain loops but which on a larger scale are connected in a loop-free way~\cite{Newman09b,Miller09,KN10b}.  For such networks one can write down exact message passing equations that operate at the higher level of the motifs and which give excellent results for problems such as structural phase transitions in networks, network spectra, and the solution of spin models~\cite{Newman09b,Miller09,KN10b,YGDM11,Newman19}.  While effective for theoretical calculations on model networks, however, this approach is of little use in practical situations.  To apply it to an arbitrary network one would first need to find a suitable decomposition of the network into motifs, and no workable method for doing this is currently known, nor even whether such a decomposition exists.

A third approach is the method known as ``generalized belief propagation,'' which has some elements in common with the motif-based approach but is derived in a different manner, from approximations to the free energy~\cite{yedidia2001generalized,yedidia2005constructing}.  This method, which is focused particularly on the solution of inference problems and related probabilistic calculations on networks, involves a hypergraph-like extension of traditional message passing that aims to calculate the joint distributions of three or more random variables at once, by contrast with the standard approach which focuses on two-variable distributions.  Generalized belief propagation was not originally intended as a method for solving problems on loopy networks but can be used in that way in certain cases.  It is, however, quite involved in practice, requiring the construction of a nested set of regions and sub-regions within the network, leading to complex sets of equations.

In this paper we take a different approach.  In the following sections we directly formulate a message passing framework that works on real-world complex networks containing many short loops by incorporating the loops themselves directly into the message passing equations.  In traditional message passing algorithms each node receives a message from each of its neighbors.  In our approach they also receive messages from nodes they share loops with.  By limiting the loops considered to a fixed maximum length, we develop a series of progressively better approximations for the solution of problems on loopy networks.  The equations become more complex as loop length increases but, as we will show, the results given by the method are already impressively accurate even at shorter lengths.

\section{Message passing}
\label{sec:mp}
Suppose we wish to calculate some value or property on the nodes of a network.  For example, we might want to calculate the probability that each person in a population contracts a disease during an epidemic or the probability that a node belongs to a particular group or community in a community detection problem.  To compute a property of a node it is in many cases sufficient to know the properties only of the node's neighbors.  For example consider the spread of a disease over the contact network of individuals in a population.  The probability that node~$i$ contracts the disease can be calculated from a knowledge of its neighbors only, since these are the only nodes from which the disease could spread.  The probability of catching the disease from a neighbor~$j$ is equal to the probability that $j$ contracts the disease times the probability that the disease is communicated from $j$ to~$i$.  If we assume that these probabilities are independent for different neighbors, then we can easily combine them to find the total probability that $i$ is infected.  Thus each node can calculate its own probability of infection in terms of those of its neighbors, a relation that can be expressed in a set of self-consistent equations.  The message passing approach involves solving these equations iteratively by guessing initial values of the probabilities for all nodes, for instance at random, and feeding those into the equations to calculate an updated set of values.  Then we repeat the process, iterating the calculation until the values converge to a fixed point, which is, by definition, a solution of the self-consistent equations.

A crucial feature of this approach is the assumption that the probabilities of different neighbors being infected are independent.  This requires that the neighbors of node~$i$ not be connected to one another, other than via~$i$ itself.  If two neighbors are directly connected by an edge, for instance, and one is infected with the disease, it makes it more likely the other is infected too, so their disease states are correlated.  An edge between two neighbors of $i$ forms a loop of length three in the network---a triangle---so another way to say the same thing is that the network must contain no triangles.  Indeed, any short path between neighbors will introduce some level of correlation, so the network cannot contain short loops of any length if the method is to give correct answers.  Correlations typically die off rapidly with path length, however, so long loops may be admissible, introducing only small inaccuracies into the calculation.  (This explains why message passing works well on locally tree-like networks, which can contain long loops but not short ones.)

As discussed in the introduction, most real-world networks contain many short loops, which creates problems for the message passing method.  One solution, as we have said, is just to ignore the loops and assume that the method works nonetheless, but in many practical situations this gives poor answers.  Here we propose an alternative approach in which loops are explicitly incorporated into the calculation.

In our approach we define a succession of approximations to the solution of the problem of interest.  In the zeroth approximation, which is equivalent to the standard message passing method described above, we assume there are no loops in our network.  Equivalently, we assume that the neighbors of a node have uncorrelated states.  In the next approximation we no longer assume that neighbors are uncorrelated.  Instead, we assume that any correlation can be accounted for by direct edges between the neighbors, which is equivalent to allowing the network to contain triangles.  In the next approximation after this, we assume that neighbor correlations can accounted for by direct edges plus paths of length 2 between neighbors.  Generally, in the $r$th approximation we assume that correlations between neighbors can be fully accounted for by paths of length $r$ and shorter.

\begin{figure}
\centering
\includegraphics[width=0.9\columnwidth]{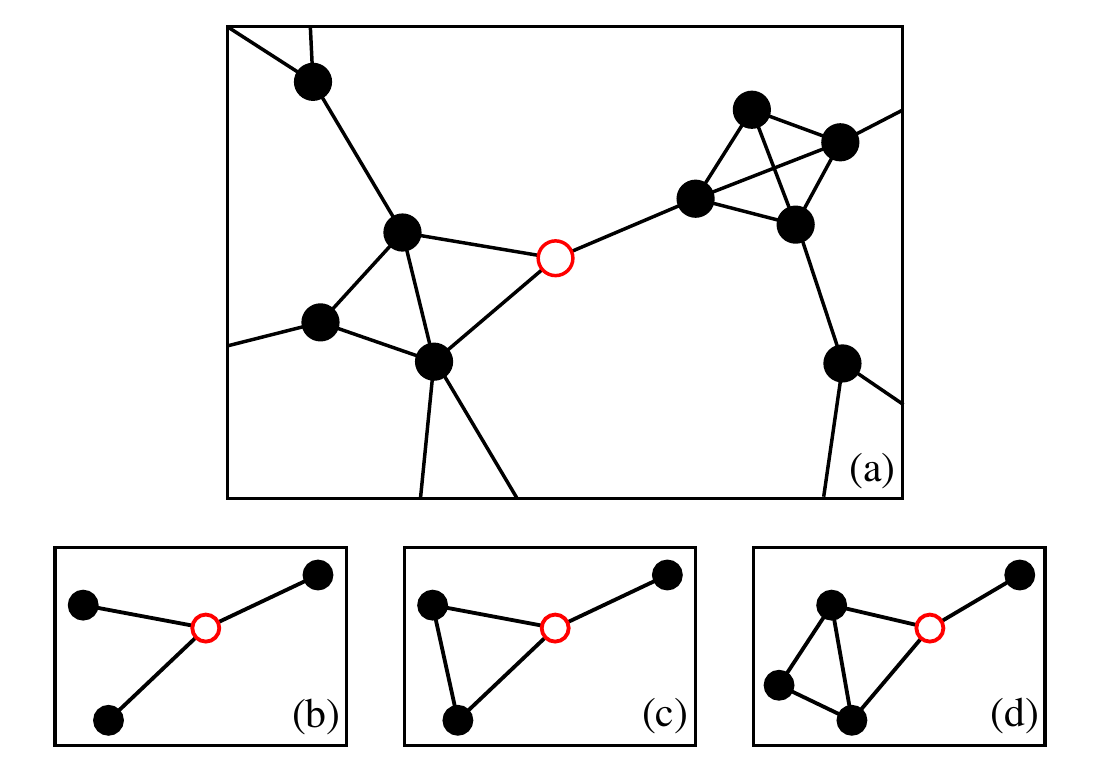}
\caption{(a)~Sketch of the focal node in our approximations (open circle) and its immediate surroundings.  (b)~In the zeroth (tree) approximation the neighborhood we consider consists of the neighbors of the focal node only.  (c)~In the first approximation we also include all length 1 paths between the neighbors.  (d)~In the second approximation we include length 2 paths as well.}
\label{fig:example_neighborhood}
\end{figure}

These successive approximations can be thought of as expressing the properties of nodes in terms of increasingly large neighborhoods and the edges they contain.  The zeroth neighborhood~$N_i^{(0)}$ of node~$i$ contains $i$'s immediate edges and nothing else.  The first neighborhood~$N_i^{(1)}$ contains node~$i$'s edges plus all length one paths between neighbors of~$i$.  The second neighborhood~$N_i^{(2)}$ contains node~$i$'s edges plus all length one and two paths between neighbors of $i$, and so forth.  Figure~\ref{fig:example_neighborhood} shows an example of how these neighborhoods are constructed.

Just as the conventional message passing algorithm is exact on trees, our algorithms will be exact on networks with short loops.  We define a \textit{primitive cycle} of length~$r$ to be a loop such that at least one node is not on a shorter loop beginning and ending at the same node.  Then our $r$th approximation is exact on networks that contain primitive cycles of length~$r+2$ and less only.  For networks that contain longer primitive cycles it will be an approximation, although as we will see it may be a good one.

\section{Message passing on networks with loops}
Our approach is best demonstrated by example.  In this section we derive message passing algorithms on loopy networks for two specific applications: the calculation of cluster sizes for bond percolation and the calculation of the spectra of sparse matrices.

\subsection{Percolation}
\label{sec:percolation}
Consider the bond percolation process on a network of $n$ nodes, where each edge is occupied independently with probability~$p$~\cite{FH63,SA92}.  Occupied edges form connected clusters and we wish to know the distribution of the sizes of these clusters and whether there exists a giant or percolating cluster that occupies a non-vanishing fraction of the network in the limit of large size.

Let us define the $r$th neighborhood~$N_i^{(r)}$ of node~$i$ as in Section~\ref{sec:mp}, then define a random variable $\Gamma_i$ for our percolation process to be the set of nodes within~$N_i^{(r)}$ that are reachable from~$i$ by traversing occupied edges only, for some specified configuration of occupied edges.  Our initial goal will be to compute the probability~$\pi_i(s)$ that node~$i$ belongs to a non-giant cluster of size~$s$.  We will do this in two stages.  First we will compute the conditional probability~$\pi_i(s|\Gamma_i)$ of belonging to a cluster of size~$s$ given the set of reachable nodes.  Then we will sum over~$\Gamma_i$ to get the full probability~$\pi_i(s)$.

Suppose that node~$i$ belongs to a cluster of size~$s$.  If our network contains no primitive cycles longer than~$r+2$, then the set of nodes~$\Gamma_i$ would become disconnected from one another were we to remove all edges in the neighborhood~$N_i^{(r)}$---the removal of these edges removes any connections within the neighborhood and there can be no connections via paths outside the neighborhood since such a path would constitute a primitive cycle of length longer than~$r+2$.  Hence the sizes~$s_j$ of the clusters to which the nodes in~$N_i^{(r)}$ would then belong must sum to~$s-1$ (the $s$th and last node being provided by~$i$ itself).  This observation allows us to write
\begin{equation}
\pi_i(s| \Gamma_i) = \!\!\sum_{\{ s_j : j \in \Gamma_i \}}
   \biggl[ \prod_{j \in \Gamma_i} \pi_{i \leftarrow j}(s_j) \biggr]
   \delta(s-1, \textstyle\sum_{j \in \Gamma_i} s_j),
\end{equation}
where $\pi_{i \leftarrow j}(s)$ is the probability that node~$j$ is in a cluster of size~$s$ once the edges in $N_i^{(r)}$ are removed.

We define a generating function for $\pi_i(s|\Gamma_i)$ as follows:
\begin{align}
H_i&(z|\Gamma_i) = \sum_{s} \pi_i(s|\Gamma_i)\,z^s  \nonumber \\
   &= \sum_{s} z^{s} \Biggl\lbrace \sum_{\{ s_j : j \in \Gamma_i \}}
   \biggl[ \prod_{j \in \Gamma_i} \pi_{i \leftarrow j}(s_j) \biggr]
   \delta(s-1, \textstyle\sum_{j \in \Gamma_i} s_j) \Biggr\rbrace \nonumber \\
   &= z \prod_{j \in \Gamma_i} \sum_{s_j} z^{s_j} \pi_{i \leftarrow j}(s_j)
    = z \prod_{j \in \Gamma_i} H_{i \leftarrow j}(z) \nonumber \\
   &= z \prod_{j \in N_i^{(r)}} \left[ H_{i \leftarrow j}(z) \right]^{w_{i j}},
\label{eq:cluster_gen_fn}
\end{align}
where in the last line we have introduced the random variable~$w_{ij}$ which takes the value~1 if $j\in\Gamma_i$ and 0 otherwise.  In other words, $w_{ij}=1$ if there is a path of occupied edges from $i$ to~$j$.

To calculate the full probability~$\pi_i(s)$ we now average $\pi_i(s|\Gamma_i)$ over sets~$\Gamma_i$ to get $\pi_i(s) = \bigl\langle \pi_i(s|\Gamma_i) \bigr\rangle_{\Gamma_i}$, where the average is weighted appropriately according the probability that each set~$\Gamma_i$ is realized, which is simply $p^k (1-p)^{m-k}$, where $p$ is the edge occupation probability as previously, $m$~is the number of network edges in the neighborhood~$N_i^{(r)}$, and $k$ is the number that are occupied.  Performing the same average on~\eqref{eq:cluster_gen_fn} gives us
\begin{align}
H_i(z) &= \sum_s \pi_i(s)\,z^s
        = \bigl\langle H_i(z|\Gamma_i) \bigr\rangle_{\Gamma_i} \nonumber \\
       &= z \Bigl\langle\!\prod_{j \in N_i^{(r)}}
          \bigl[ H_{i \leftarrow j}(z) \bigr]^{w_{i j}} \Bigr\rangle_{\Gamma_i}
        = z G_i\bigl( \vec{H}_{i \leftarrow}(z) \bigr),
\label{eq:cluster_gen_fn2}
\end{align}
where $G_i(\vec{y})$ is a generating function for~$w_{i j}$:
\begin{equation}
G_i(\vec{y}) =  \Bigl\langle\!\prod_{j \in N_i^{(r)}}\!y_j^{w_{ij}} \Bigr\rangle_{\Gamma_i},
\label{eq:defsGi}
\end{equation}
and $\vec{H_{i \leftarrow}}(z)$ is the vector with elements $H_{i \leftarrow j}(z)$ for nodes $j$ in~$N_i^{(r)}$.

To complete the calculation we still need to evaluate the~$H_{i \leftarrow j}(z)$, whose computation follows the same logic as for~$H_i(z)$, the only difference being that in considering the neighborhood~$N_j^{(r)}$ of node~$j$ we must remove the entire neighborhood of $i$~first, as described above.  We can then derive a generating function
\begin{equation}
H_{i \leftarrow j}(z| \Gamma_{j \setminus i}) = z\!\!\prod_{k \in N_{j \setminus i}^{(r)}} \bigl[ H_{j \leftarrow k}(z) \bigr]^{w_{j k}},
\end{equation}
where $N_{j \setminus i}^{(r)}$ is $N_j^{(r)}$ with~$N_i^{(r)}$ removed and $\Gamma_{j \setminus i}$ is the set of nodes that can be reached from $j$ by traversing occupied edges in~$N_{j \setminus i}^{(r)}$, for some realization of edge occupancies.  Then, averaging over~$\Gamma_{j \setminus i}$, we obtain
\begin{equation}
H_{i \leftarrow j}(z) = z G_{i \leftarrow j}\bigl( \vec{H}_{j \leftarrow}(z) \bigr).
\label{eq:self_consistent_message_eq}
\end{equation}
If we can solve this equation self-consistently for~$\vec{H}_{j \leftarrow}(z)$, we can substitute the solution into Eq.~\eqref{eq:cluster_gen_fn2} to compute the full cluster size generating function.  The message passing method solves~\eqref{eq:self_consistent_message_eq} by simple iteration: we choose suitable starting values, for instance at random, and iterate the equations to convergence.

From the cluster size generating function we can calculate a range of quantities of interest.  For example, the probability that node~$i$ belongs to a small cluster (of any size) is $H_i(1) = \sum_s \pi_i(s)$.  If it does not belong to a small cluster then necessarily it is in the percolating cluster and hence the expected fraction~$S$ of the network taken up by the percolating cluster is
\begin{equation}
S = 1 - \frac{1}{n} \sum_i H_i(1).
\end{equation}
Similarly, the average value of~$s_i$ is
\begin{align}
\langle s_i \rangle &= \sum_s s \pi_i(s) = H_i'(1) \nonumber \\
  &= H_i(1) + \sum_{j \in N_i^{(r)}} \! H_{i \leftarrow j}'(1)\,
     \partial_j G_i( \vec{H}_{i \leftarrow}),
\end{align}
where $H'$ is the derivative of~$H$ and $\partial_j G_i$ is the partial derivative of $G_i$ with respect to its $j$th argument.  $H_{i \leftarrow j}'(1)$ can be found by differentiating Eq.~\eqref{eq:self_consistent_message_eq} and setting~$z=1$ to give the self-consistent equation
\begin{align}
H_{i \leftarrow j}'(1) &=  H_{i \leftarrow j}(1) + \sum_{k \in N_{j \setminus i}^{(r)}} H_{j \leftarrow k}'(1)\,\partial_k G_{i \leftarrow j }\left( \vec{H}_{j \leftarrow }\right).
\label{eq:H_deriv}
\end{align}

While these equations are straightforward in principle, implementing them in practice presents some additional technical challenges.  Computing the generating functions~$G_i(\vec{y})$ and $G_{i \leftarrow j}(\vec{y})$ can be demanding, since it requires us to perform an average over the occupancy configurations of all edges within the neighborhoods~$N_i^{(r)}$ and~$N_{j\setminus i}^{(r)}$, and the number of configurations increases exponentially with neighborhood size.  For small neighborhoods, such as those on low-dimensional lattices, it is feasible to average exhaustively, but for many complex networks this is not possible.  In such cases we instead approximate the average by Monte Carlo sampling of configurations---see Appendix~A for details.  A nice feature of the Monte Carlo procedure, as described in the appendix, is that the samples only need be taken only once for the entire calculation and can then be reused on successive iterations of the message passing process.

\begin{figure}
\centering
\includegraphics[width=\columnwidth]{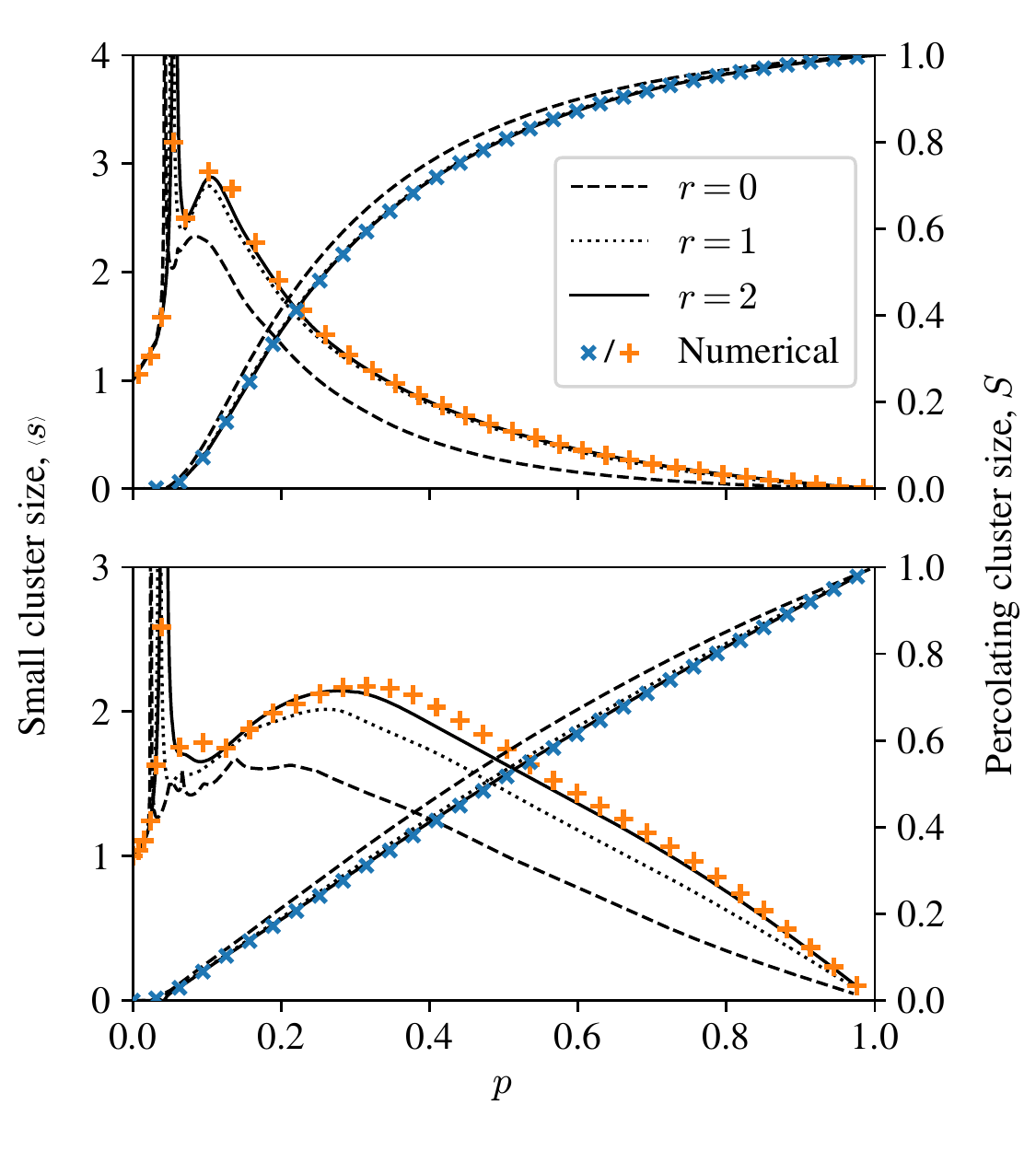}
\caption{Percolating cluster size ($\times$~symbols) and average cluster size (+~symbols) for two real-world networks.  Top: the largest component of a coauthorship network of $13\,861$ scientists~\cite{Newman01a}.  Bottom: a network of $10\,680$ users of the PGP encryption software~\cite{BPDA04b}.}
\label{fig:percolation}
\end{figure}

In practice the method gives excellent results.  We show example applications to two real-world networks in Fig.~\ref{fig:percolation}.  The first is a social network of coauthorship relations between scientists in the field of condensed matter physics~\cite{Newman01a}.  The second is a network of trust relations between users of the PGP encryption software~\cite{BPDA04b}.  Both have a high density of short loops.  For each network the figure shows, as a function of~$p$, several different estimates of both the average size~$\langle s\rangle$ of the small clusters and the size~$S$ of the percolating cluster as a fraction of~$n$.  First we show an estimate made using standard message passing (dashed line)---the $r=0$ approximation in our nomenclature---which ignores loops and is expected to give poor results.  Second, we show the next two approximations in our series, those for $r=1$ and $r=2$ (dotted and solid lines respectively).  For these approximations we estimate $G_i(\vec{y})$ and $G_{i \leftarrow j}(\vec{y})$ by Monte Carlo sampling as described above.
We use only eight samples for each node~$i$ but the results are nonetheless impressively accurate.  Third, we show for comparison a direct numerical estimate of the quantities in question made by conventional simulation of the percolation process.

For both networks we see the same pattern.  The traditional message passing method fares poorly, as expected, giving estimates that are substantially in disagreement with the simulation results, particularly for the calculations of average cluster size.  The $r=1$ approximation, on the other hand, does much better, agreeing almost perfectly with the numerical results in most cases, although less well for the average cluster size results in the PGP network.  The $r=2$ approximation does even better, agreeing closely with the numerical results for all measures on all networks.  In these examples at least, it appears that the $r=2$ method gives accurate results for bond percolation, where standard message passing fails.

The message passing algorithm is relatively fast.  For $r \le 1$ each node receives a message from each neighbor on each iteration, and so on a network with mean degree $c$ there are $cn$ messages passed per iteration.  For $r \ge 2$ the number of messages depends on the network structure.  On trees the number of messages remains unchanged at $cn$ as $r$ increases but on networks with loops it grows and for large numbers of loops it can grow exponentially.  In the common case of sparse networks, however, the size of the neighborhoods does not grow with~$n$, so each iteration is linear in~$n$ for fixed~$r$.  It is not known in general how many iterations are needed for message passing methods to reach convergence, but elementary heuristic arguments suggest the number should be on the order of the diameter of the network, which is typically~$\Ord(\log n)$~\cite{WS98}.  Thus we expect overall running time to be $\Ord(n \log n)$ for sparse networks at fixed~$r$.

This makes the algorithm quite efficient, although direct numerical simulations of percolation run comparably fast, so the message passing approach does not offer a speed advantage over traditional approaches.  However, the two approaches are calculating different things.  Traditional simulations of percolation perform a calculation for one particular realization of bond occupancies.  If one wants average values over many realizations one must perform the average explicitly, repeating the whole simulation for each realization.  The message passing approach, on the other hand, computes the average over realizations in a single calculation and no repetition is necessary.

In the next section we demonstrate another example application of our methods, to the calculation of the spectrum of a sparse matrix, where traditional and message passing calculations differ substantially in their running time, the message passing approach being much faster, making calculations possible for large systems whose spectra cannot be computed in any reasonable amount of time by traditional means.

\subsection{Matrix spectra}
\label{sec:spectral_density}
For our second example application we show how the message passing method can be used to compute the eigenvalue spectrum of a sparse symmetric matrix.  Any $n \times n$ symmetric matrix can be thought of as an undirected weighted network on $n$ nodes and we can use this equivalence to apply our message passing method to such matrices.

The spectral density of a symmetric matrix $\mat{A}$ is the quantity
\begin{equation}
\rho(x) = \frac{1}{n} \sum_{i=1}^{n} \delta(x - \lambda_i),
\label{eq:density1}
\end{equation}
where $\lambda_i$ is the $i$th eigenvalue of $\mat{A}$, and $\delta(x)$ is the Dirac delta function.  Following standard arguments~\cite{NN13}, we can show that the spectral density is equal to the imaginary part of the complex function
\begin{align}
\rho(z) &= -\frac{1}{n \pi} \sum_{i=1}^{n} \frac{1}{z-\lambda_i}
   = -\frac{1}{n \pi} \Tr(z \mat{I} - \mat{A} )^{-1} \nonumber \\
	&= -\frac{1}{n \pi z} \sum_{s=0}^{\infty}\sum_{i=1}^{n} \frac{X_i^s}{z^s},
\label{eq:density2}
\end{align}
where $X_i^s = [\mat{A}^s]_{ii}$ is the $i$th diagonal element of~$\mat{A}^s$, and $z = x + i\eta$ and we take the limit as $\eta\to0$ from above.  The imaginary part~$\eta$ acts as a resolution parameter that broadens the delta-function peaks in~\eqref{eq:density1} by an amount roughly equal to its value.

The quantities $X_i^s =[\mat{A}^s]_{i i}$ can be related to sums over closed walks in the equivalent network.  If we consider the ``weight'' of a walk to be the product of the matrix elements on the edges it traverses, then $X_i^s$ is the sum of the weights of all closed walks of length $s$ that start and end at node~$i$.

A closed walk from~$i$ need not visit $i$ only at its start and end.  It can return to~$i$ any (positive) number of times over the course of the walk.  The simplest case, where it returns just once at the end of the walk, we will call an \emph{excursion}.  A more general closed walk that returns to node~$i$ exactly $m$ times can be thought of as a succession of $m$ excursions.  Such a walk will have length~$s$ if those $m$ excursions have lengths $s_1, \dots, s_m$ with $\sum_{u=1}^m s_u = s$.

With this in mind, let $Y_i^s$ be the sum of the weights of all \emph{excursions} of length~$s$ that start and end at node~$i$.  Then the sum~$X_i^s$ over \emph{closed walks} of length~$s$ can be written in terms of~$Y_i^s$~as
\begin{equation}
X_i^s = \sum_{m=0}^{\infty} \left[ \sum_{s_1=1}^{\infty} \dots \sum_{s_m=1}^{\infty}  \delta \bigl(s, {\textstyle\sum_{u=1}^{m}} s_u \bigr) \prod_{u=1}^{m} Y_i^{s_u} \right].
\label{eq:X_i^s}
\end{equation}
Substituting this expression into Eq.~\eqref{eq:density2} we get
\begin{equation}
\rho(z) = -\frac{1}{n \pi z} \sum_{i=1}^{n} \sum_{m=0}^{\infty}\,\prod_{u=1}^{m} \Biggl[ \sum_{s=1}^{\infty} \frac{ Y_i^s }{z^{s}} \Biggr],
\label{eq:rhoy}
\end{equation}
and, defining the function
\begin{equation}
H_i(z) = \sum_{s=1}^{\infty} \frac{Y_i^s}{z^{s-1}},
\label{eq:closed_walk_gen_fun}
\end{equation}
we find that
\begin{equation}
\rho(z) = -\frac{1}{n \pi z} \sum_{i=1}^{n} \sum_{m=0}^{\infty} \biggl[ \frac{H_i(z)}{z} \biggr]^m 
  = -\frac{1}{n \pi} \sum_{i=1}^{n} \frac{1}{z-H_i(z)}.
\label{eq:final_rho}
\end{equation}
Thus, if we can calculate~$H_i(z)$ then we can calculate~$\rho(z)$.  This we do by message passing as follows.

\begin{figure}
\centering
\includegraphics[width=0.65\columnwidth]{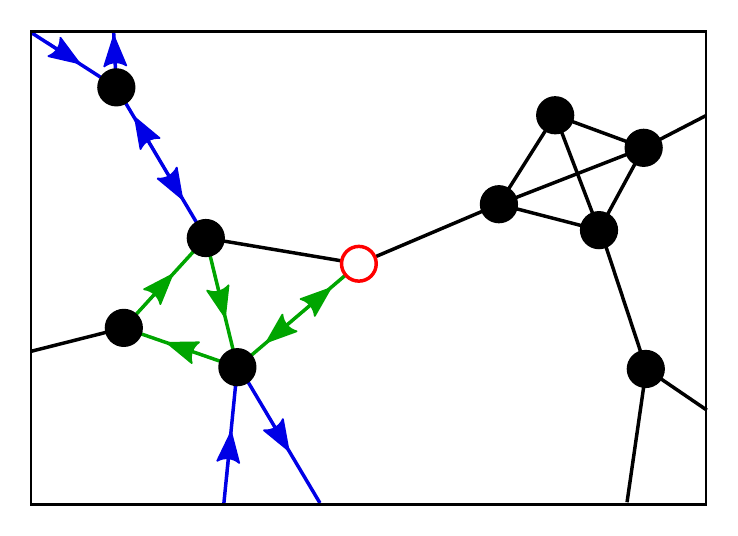}
\caption{An example excursion from the central node (open circle). The excursion is equivalent to an excursion inside the neighborhood, shown with green arrows,  plus closed walks to regions outside of the neighborhood, shown in blue.}
\label{fig:example_walk}
\end{figure}

Consider the neighborhood~$N_i^{(r)}$ around~$i$ as defined in Section~\ref{sec:mp}.  If there are no primitive cycles of length longer than~$r+2$ in our network then all loops starting at~$i$ are already included within the neighborhood, which means that any excursion from $i$ takes the form of an excursion~$w$ within the neighborhood plus some number of additional closed walks outside the neighborhood that each start at one of the nodes in~$w$ and return some time later to the same node---see Fig. \ref{fig:example_walk}.  The additional walks must necessarily return to the same node they started at since if they did not they would complete a closed walk from~$i$ that lies outside the neighborhood, of which by hypothesis there are none.

Let the length of the excursion~$w$ be~$l+1$, meaning that it visits $l$ nodes~$j_1\dots j_l$ (not necessarily distinct) within the neighborhood other than the starting node~$i$, and let $s_j$ be the length of the external closed walk (if any) that starts at node~$j$, or zero if there is no such walk.  The total length of the complete excursion from~$i$ will then be $l+1+\sum_{j \in w} s_j$ and the sum of the weights of all excursions of length~$s$ with $w$ as their foundation will be
\begin{equation}
|w|\!\!\sum_{ \{ s_j : j \in w \} } \!\!\delta\bigl(s,l+1+\textstyle{\sum_{j \in w} s_j}\bigr)\displaystyle \prod_{j \in w} X^{s_j}_{i \leftarrow j},
\label{eq:excursions_length_s}
\end{equation}
where $|w|$ is the weight of $w$ itself and $X^s_{i \leftarrow j}$ is the sum of weights of length-$s$ walks from node~$j$ if the neighborhood~$N_i^{(r)}$ is removed from the network.  By a similar argument to the one that led to Eq.~\eqref{eq:X_i^s}, we can express $X^s_{i \leftarrow j}$ in terms of the sum $Y^s_{i \leftarrow j}$ of excursions from~$j$ thus:
\begin{equation}
X_{i\leftarrow j}^s = \sum_{m=0}^{\infty} \left[ \sum_{s_1=1}^{\infty} \dots \sum_{s_m=1}^{\infty}  \delta \bigl(s, {\textstyle\sum_{u=1}^m} s_u \bigr) \prod_{u=1}^{m} Y_{i\leftarrow j}^{s_u} \right].
\label{eq:X_ij^s}
\end{equation}
And the quantity~$Y_i^s$ appearing in Eq.~\eqref{eq:closed_walk_gen_fun} can be calculated by summing~\eqref{eq:excursions_length_s} first over the set~$W_i^l$ of excursions of length $l+1$ in the neighborhood of~$i$ and then over~$l$, thus:
\begin{equation}
Y_i^s = \sum_{l=0}^\infty \sum_{w \in W_i^l} |w|\!\!\! \sum_{\{ s_j : j \in w \}} \!\!\! \delta\bigl(s,l+1+\textstyle{\sum_{j \in w} s_j}\bigr)\displaystyle \prod_{j \in w} X^{s_j}_{i \leftarrow j}.
\label{eq:Y_i^s}
\end{equation}
Combining Eqs.~\eqref{eq:X_ij^s} and~\eqref{eq:Y_i^s} and substituting into~\eqref{eq:closed_walk_gen_fun} we now find that
\begin{align}
H_i(z) &= \sum_{l=0}^{\infty} \frac{1}{z^l}\!\!\sum_{w \in W_i^l} |w| \prod_{j \in w}\, \sum_{m=0}^{\infty}\,\prod_{k=1}^{m}\,\sum_{s=1}^{\infty} \frac{Y^{s}_{i \leftarrow j}}{z^{s}} \nonumber \\
  &= \sum_{w \in W_i} \!|w| \prod_{j \in w} \frac{1}{z - H_{i \leftarrow j}(z)}\,,
\label{eq:Hi}
\end{align}
where $W_i$ is now the complete set of excursions of all lengths in the neighborhood of~$i$ and we have defined
\begin{equation}
H_{i \leftarrow j}(z) = \sum_{s=1}^{\infty} \frac{Y^{s}_{i \leftarrow j }}{z^{s-1}}.
\end{equation}
Following an analogous line of argument for this function we can show similarly that
\begin{equation}
H_{i \leftarrow j}(z) = \sum_{w \in W_{j \setminus i}} \!\! |w|
  \prod_{k \in w} \frac{1}{z-H_{j \leftarrow k}(z)}.
\label{eq:spectral_message2}
\end{equation}

Equation~\eqref{eq:spectral_message2} defines our message passing equations for the spectral density.  By iterating these equations to convergence from suitable starting values we can solve for the values of the messages~$H_{i \leftarrow j}(z)$, then substitute into Eqs.~\eqref{eq:Hi} and~\eqref{eq:final_rho} to get the spectral density itself.

As with our percolation example, the utility of this approach relies on our having an efficient method for evaluating the sum in Eq.~\eqref{eq:spectral_message2}.  Fortunately there is such a method, as follows.  Let $v_{i \leftarrow j, k} = A_{j k}$ if nodes $j$~and~$k$ are directly connected in $N_{j \setminus i}^{(r)}$ and $0$ otherwise.  Further, let $\mathbf{A}^{i \leftarrow j}$ be the adjacency matrix of the neighborhood of $j$ with the neighborhood of $i$ removed, such that
\begin{equation}
A_{kl}^{i \leftarrow j} = \biggl\lbrace\begin{array}{ll}
	A_{k l} & \quad\mbox{for $k,l\neq j$ and edge $(k,l)\in N_{j \setminus i}^{(r)},$} \\
  0 & \quad\mbox{otherwise,}
\end{array}
\end{equation}
and let $\mathbf{D}^{i \leftarrow j}(z)$ be the diagonal matrix with entries $D_{kk}^{i \leftarrow j} = z-H_{j \leftarrow k}(z)$.  Equation~\eqref{eq:spectral_message2} can then be written as
\begin{equation}
H_{i \leftarrow j}(z) = \vec{v}_{i \leftarrow j}^T \bigl( \mat{D}^{i \leftarrow j} - \mat{A}^{i \leftarrow j} \bigr)^{-1} \vec{v}_{i \leftarrow j} + A_{jj}.
\label{eq:matrix_method}
\end{equation}
Since the matrices in this equation are the size of the neighborhood, each message update requires us to invert only a small matrix, which gives us a linear-time algorithm for each iteration of the message passing equations and an overall running time of $\Ord(n \log n)$ for sparse networks with fixed degree distributions, or for the equivalent sparse matrices.

\begin{figure}
\centering
\includegraphics[width=\columnwidth]{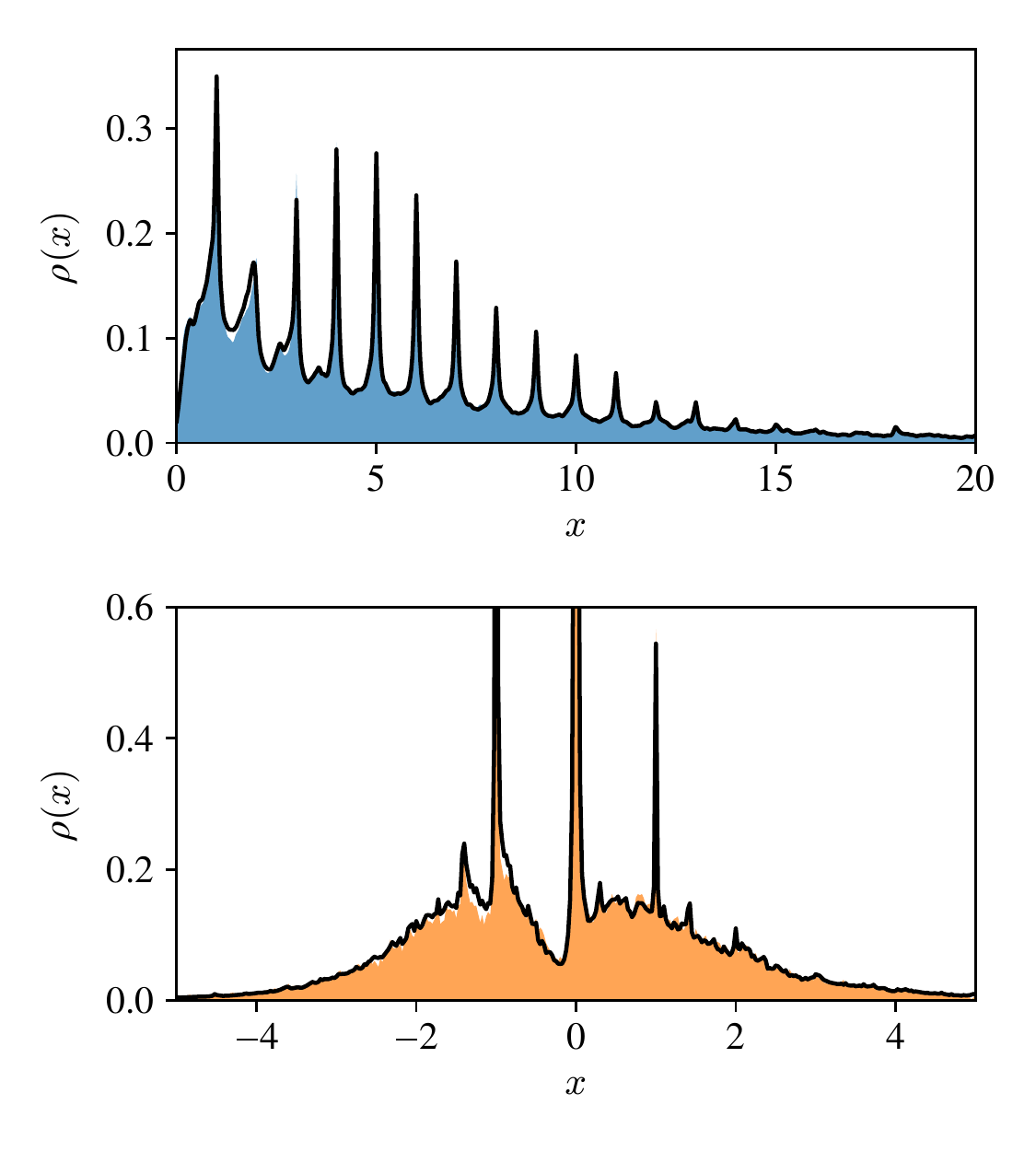}
\caption{Matrix spectra for the same two networks that were used in Fig.~\ref{fig:percolation}.  Top: the spectrum of the graph Laplacian of the coauthorship network.  Bottom: the spectrum of the adjacency matrix of the PGP network.  The shaded areas show the spectral density calculated by direct numerical diagonalization.  The black lines show the $r=1$ message-passing approximation.  The broadening parameter~$\eta$ was set to $0.05$ in the top panel and $0.01$ in the bottom panel.}
\label{fig:spectra}
\end{figure}

As an example of the application of this method, we show in Fig.~\ref{fig:spectra} spectra for the same two real-world networks that we used in Fig.~\ref{fig:percolation}.  To demonstrate the flexibility of the method we calculate different spectra in the two cases: for the coauthorship network we calculate the spectrum of the graph Laplacian; for the PGP network we calculate the adjacency matrix spectrum.

For each network we calculate the spectral density using our message passing method with $r=1$, shown as the black curve in each panel.  We also calculate the full set of eigenvalues of each network directly using traditional numerical methods and substitute the results into Eq.~\eqref{eq:density2} to compute the spectral density, shown as the shaded area in each panel.  As the figure shows, the agreement between the two is excellent.  There are a few regions where small differences are visible but in general they agree closely.  Extending the calculation to the next ($r=2$) approximation gives a modest further improvement in the results.

The $\Ord(n \log n)$ running time of the message passing algorithm significantly outstrips that of traditional numerical diagonalization.  Complete spectra are normally calculated using the QR algorithm, which runs in time~$\Ord(n^3)$ and is consequently much slower as system size becomes large.  The Lanczos algorithm runs in $\Ord(n \log n)$ time on sparse matrices, but typically gives only a few leading eigenvalues and not a complete spectrum.  The kernel polynomial method~\cite{WWAF06} is capable of computing complete spectra for sparse matrices, but requires Monte Carlo evaluation of the traces of large matrix powers which has slow convergence and is always only approximate, even in cases where our method gives exact results.

This opens up the possibility of using our approach to calculate the spectral density of networks and matrices significantly larger than those that can be tackled by traditional means.  As an example, we have used the message passing method to compute the spectral density of one network with $317\,080$ nodes.  This is significantly larger than the largest systems that can be diagonalized using the QR algorithm, which on current (non-parallel) commodity hardware is limited to a few tens of thousands of nodes in practical running times.

\section{Conclusions}
In this paper we have described a new class of message passing methods for performing calculations on networks that contain short loops, a situation in which traditional message passing often gives poor results or may fail to converge entirely.  We derive message passing equations that account for the effects of loops up to a fixed length that we choose, so that calculations are exact on networks with no loops longer than this.  In practice, however, we achieve excellent results on real-world networks by accounting for loops up to length three or four only.

We have demonstrated our approach with two example applications, one to the calculation of bond percolation properties of networks and the other to the calculation of the spectra of sparse matrices.  In the first case we develop message passing equations for the size of the percolating cluster and the average size of small clusters and find that these give good results, even on networks with an extremely high density of short loops.  For the calculation of matrix spectra, we develop a message passing algorithm for the spectral density that gives results in good agreement with traditional numerical diagonalization but in much shorter running times.  Where traditional methods are limited to matrices with at most a few tens of thousands of rows and columns, our method can be applied to cases with hundreds of thousands at least.

There are a number of potential directions for further work on this topic.  Chief among them is the application of the method to other classes of problems, such as epidemic models, graph coloring, or satisfiability problems.  Many extensions of the calculations in this paper are also possible, including the inclusion of longer primitive cycles in the message passing equations, development of more efficient algorithms for very large systems, and applications to individual examples of interest such as the computation of spectra for very large graphs.  These possibilities, however, we leave for future research.

\begin{acknowledgments}
The authors thank Cristopher Moore for useful conversations.  This work was funded in part by the US National Science Foundation under grant DMS--1710848.
\end{acknowledgments}

\begin{center}
\rule{4cm}{0.5pt}
\end{center}

\appendix
\section{Monte Carlo algorithm for $G_i(\vec{y})$}
In the message passing construction for bond percolation described in Section~\ref{sec:percolation}, the quantity~$G_i(\vec{y})$ is a generating function encoding the probability that we can reach nodes in the neighborhood~$N_i^{(r)}$ of a given node~$i$ by following occupied edges.  It is defined by
\begin{equation}
G_i(\vec{y}) = \Bigl\langle \prod_{j \in N_i^{(r)}} y_j^{w_{ij}} \Bigr\rangle_{\Gamma_i}\,,
\end{equation}
where $w_{ij}$ is a binary (zero/one) random variable indicating whether node~$j$ is reachable from node~$i$ and the average is performed over all sets $\Gamma_i$~of reachable nodes in the neighborhood.  As discussed in Section~\ref{sec:percolation}, the number of possible such sets can become large as the size of the neighborhood grows, making exhaustive averages difficult to perform numerically.  For larger neighborhoods, therefore, we employ a Monte Carlo averaging scheme as follows.

Suppose that node~$i$ has degree~$k_i$ and that there are $k_i+M$ edges in the neighborhood~$N_i^{(r)}$, with $k_i$ of them directly connected to~$i$ and $M$ additional edges that complete cycles between $i$'s neighbors.  For locally tree-like networks there are no cycles and $M=0$, but in general $M \geq 0$.  Let $G_i(\vec{y} | m)$ be the value of $G_i(\vec{y})$ when exactly $m$ of the $M$ additional edges are occupied, which happens with probability~${M \choose m} p^m(1-p)^{M-m}$.  Then we can write $G_i(\vec{y})$ itself in the form
\begin{equation}
G_i(\vec{y}) = \sum_{m=0}^M G_i(\vec{y} | m) {M \choose m} p^m(1-p)^{M-m}.
\label{eq:gialg}
\end{equation}

Our algorithm works by making a Monte Carlo estimate of $G_i(\vec{y} | m)$ using a version of the algorithm of Newman and Ziff~\cite{NZ00} and then applying Eq.~\eqref{eq:gialg}.  The basic idea is to occupy edges one by one and keep track of the connected percolation clusters using an efficient union-find data structure based on pointers~\cite{NZ00}.  Using this data structure the algorithm is able to determine whether two nodes belong to the same cluster, or to join two clusters together, in (very nearly) $\Ord(1)$ time.  To compute $G_i(\vec{y} | m)$ itself, the algorithm maintains a record of two quantities for each cluster, a real value~$x$ and a probability~$q$.  In detail the algorithm works as follows.

The clusters we consider are the sets of nodes other than~$i$ that are connected via occupied edges in $N_i{(r)}$ but not via node~$i$ itself, i.e.,~via the $M$ additional edges mentioned above.  Initially none of the $M$ edges is occupied and each node is a cluster in its own right.  For each of these one-node clusters~$j$ we assign $x_j=y_j$ and we set $q_j=1-p$ if node~$j$ is a direct neighbor of~$i$ or $q_j=1$ otherwise.  We also compute the quantity
\begin{equation}
u_0 = \prod_{j} \left( q_j + \left(1-q_j\right)x_j \right).
\end{equation}

Now we occupy the $M$ edges one by one in random order.  Let $j_1$ and $j_2$ be the nodes at the ends of the $m$th edge added.  If $j_1$ and $j_2$ are already part of the same cluster before the edge is added (which, as we have said, we can determine in time~$\Ord(1)$), then we set
\begin{equation}
u_m \leftarrow u_{m-1}.
\end{equation}
Otherwise, if $j_1$ and $j_2$ are in different clusters $r$ and~$s$, then the addition of the $m$th edge joins $r$ and $s$ together (which again we can achieve in~$\Ord(1)$ time) to make a larger cluster which, without loss of generality, we will label~$r$.  At the same time we also set
\begin{align}
u_m &\leftarrow \frac{u_{m-1}}{ \big( q_r + \left(1-q_r\right)x_r \big) \big(q_s + (1-q_s)x_s \big)}, \\
x_r &\leftarrow x_r x_s, \\
q_r &\leftarrow q_r q_s, \\
u_m &\leftarrow u_{m} \left(q_r + (1-q_r)x_r \right).
\end{align}
After all $M$ edges have been occupied, the $M+1$ quantities~$u_m$ with $m=0\ldots M$ give us an estimate of~$G_i(\vec{y} | m)$ and $G_i(\vec{y})$ can be calculated from Eq.~\eqref{eq:gialg} as
\begin{equation}
G_i(\vec{y}) \simeq \sum_{m=0}^\infty u_m {M\choose m} p^m (1-p)^{M-m}.
\label{eq:gialg2}
\end{equation}
The computation for $G_{i \leftarrow j}(\vec{y})$ is identical except for the replacement of the neighborhood by~$N_{j \setminus i}^{(r)}$.  Finally, we average the results over repeated runs of the algorithm to get our estimate of the generating functions.  As mentioned in Section~\ref{sec:percolation}, we find surprisingly good results with averages over a relatively small number of runs---we used just eight runs for each neighborhood to generate the results shown in Fig.~\ref{fig:percolation}.

Note that the sequence of edges added and cluster joins performed does not depend on the values of either~$\vec{y}$ or~$p$, which means we can use the same sequence to calculate~$G_i(\vec{y})$ for many different~$\vec{y}$ and~$p$.  We can also use the same sequence on successive iterations of the message passing process, which has the benefit of removing any statistical fluctuations between iterations and is useful when estimating convergence of the message passing process, which can otherwise be difficult to do.

\end{document}